\documentstyle[preprint,eqsecnum,aps,epsfig,floats]{revtex}
\tighten
\begin{document}
\draft
\preprint{IRPHE/1997-LRV}

\title{Hamiltonian Dynamics and the Phase Transition of the XY Model}

\author{Xavier Leoncini, and Alberto D. Verga,%
\thanks{
        e-mail: verga@marius.univ-mrs.fr}}
\address{
Institut de Recherche sur les Ph\'enom\`enes Hors Equilibre,%
\thanks{
        Unit\'e Mixte de Recherche 6594, Centre National de la
        Recherche Scientifique, Universit\'es d'Aix-Marseille I et
        II.}\\
12, avenue G\'en\'eral Leclerc, F-13003 Marseille, France.}

\author{and Stefano Ruffo}
\address{
Dipartimento di Energetica ``S. Stecco",\\
Universit\`a degli Studi di Firenze, INFN and INFM,\\
Via di S.Marta, 3 50139 Firenze, Italy.}

\date{\today}
\maketitle

\begin{abstract}
A Hamiltonian dynamics is defined for the XY model by adding a
kinetic energy term. Thermodynamical properties (total energy,
magnetization, vorticity) derived from microcanonical simulations of
this model are found to be in agreement with canonical Monte-Carlo
results in the explored temperature region. The behavior of the
magnetization and the energy as functions of the temperature are
thoroughly investigated, taking into account finite size effects. By
representing the spin field as a superposition of random phased
waves, we derive a nonlinear dispersion relation whose solutions
allow the computation of thermodynamical quantities, which agree
quantitatively with those obtained in numerical experiments, up to
temperatures close to the transition. At low temperatures the
propagation of phonons is the dominant phenomenon, while above the
phase transition the system splits into ordered domains separated by
interfaces populated by topological defects. In the high temperature
phase, spins rotate, and an analogy with an Ising-like system can be
established, leading to a theoretical prediction of the critical
temperature $T_{KT}\approx 0.855$.
\end{abstract}

\section{Introduction}
\label{Sec:Introduction}

The two dimensional XY model, also known as the planar spin model,
presents many interesting behaviors. Despite the presence of a
continuous symmetry group, a particular form of phase transition
exists \cite{Itz89,LeB88}, which can be characterized by the change
in the behavior of the correlation functions. In the low temperature
phase these latter have power law decay, showing that the system is
in a long range order state; while they decay exponentially at high
temperatures, the long range order is broken, even though
thermodynamic quantities remain smooth across the
transition~\cite{Ber71}. These observations have been interpreted by
Kosterlitz and Thouless \cite{KT73} using an analogy with the
transition of a Coulomb gas from a dielectric phase, where charges
are bounded into dipoles, to a plasma (conducting) phase where
temperature fluctuations destroy the dipoles, and the charges become
free. In the XY system the charges are replaced by topological
excitations called vortices.

From the analytical point of view, beyond the spin-wave approximation
\cite{Ber73}, and the Villain model \cite{Vil80}, the use of
renormalization group in the critical region has been the main issue
\cite{Kos74,Min87}. In order to confirm the analytical results and to
satisfy the need for a better understanding of the transition, many
numerical studies have been performed. Different dynamics, like
Monte-Carlo \cite{Tob79} or Langevin \cite{Yur93}, have been
introduced. The system has been confirmed to be dominated by spin
wave excitations in the low temperatures region. The transition
region is then determined using the information on the correlation
functions, and is interpreted in terms of dipoles unbinding. However,
there is a number of observations in the literature that seems to
complicate this simple picture of the phase transition mechanism.
Among the different results those worth mentioning are: the
visualization of the vortex distribution \cite{Tob79}, the presence
of domains delimited by topological defects \cite{Yur93}, the precise
determination of the transition temperature $T_{KT}\approx 0.89$
\cite{Ols94}, much lower than the transition temperature of the
Villain model. Results are also found on the interaction between
vortices \cite{Yur93}, or the vortex interaction energy \cite{Web91}.

Although the basic mechanism of the Kosterlitz-Thouless transition,
in terms of the breaking of vortex dipoles associated to the
emergence of a disordered state, is well understood, the observation
of the spatial distribution of defects, which is not uniform (defects
tend to appear organized into clusters at temperatures slighty larger
than the transition temperature), and the presence of large ordered
domains where the spins are almost parallel, seem to indicate that
the physics of the phase transition is not exhausted by this
unbinding process but that some kind of partial local order is
present even beyond the transition temperature. The investigation of
the system properties near the transition is one of the points
addressed in this paper.

On the other hand, the question whether statistical physics is able
to describe over a wide range of temperatures the behavior of a
classical Hamiltonian system with many degrees of freedom, still
remains an open one (see e.g. the literature on the Fermi-Pasta-Ulam
model quoted in \cite{Ford92}). Therefore, it seemed interesting to
consider the XY model from the point of view of Hamiltonian dynamical
systems by adding a kinetic energy term to the XY Hamiltonian. Such
kind of approach has proved to be fruitful of interesting information
on typical relaxation time scales and collective behaviors in the 1D
case \cite{Esc94} and in the mean field approximation \cite{Ant96}.
More recently Lyapunov exponents have been computed, confirming the
presence of long relaxation times to equilibrium both in the very low
and in the very high temperature limits in one dimension \cite{Cas96} and the
presence of a change in slope of the maximal Lyapunov exponent vs
the energy near the transition temperature in 2D \cite{Pos96} (see
also Ref.~\cite{But80} for a preliminary study of this latter
phenomenon).

The present model is one of coupled rotators (spins) sitting on the
sites of a square lattice interacting with near neighbors, whose
statistical properties are described by the microcanonical ensemble,
the total energy being set by the initial conditions. Convergence to
a Gibbsian equilibrium distribution is not given for granted in the
whole temperature range on accessible time scales. Therefore, a
particular attention must be devoted to the temporal behavior of the
different quantities which characterize macroscopic thermodynamical
properties. One notices, however, that the spatial topology is
unchanged and that the existence of a continuous symmetry group of
rotation is also maintained in the dynamics. This implies that a
Kosterlitz-Thouless type transition must be observed (in the
thermodynamic limit) with  typically strong finite size effects,
like the existence of a non-zero magnetization \cite{Ber73,Bra94}.

In this paper we concentrate on a study of those properties of the
dynamics of the XY model which reproduce equilibrium features,
postponing to a future work the study of non equilibrium effects. We
 anticipate that we are able to reproduce most of the equilibrium
behaviors of macroscopic quantities, which makes microcanonical
dynamics competitive with canonical Monte Carlo, since we have not
here to extract random numbers, stochasticity being supplied by the
intrinsic chaoticity of the model.

The fact that we actually deal with a dynamical system, allows us to
develop an original analytical approach to the study of the
thermodynamics, which is based on the approximate solution of the
equations of motion. This method is based upon the ergodic properties
of the dynamics and on the separation of temporal scales present in
the spin motion.

In Section \ref{SecModel} we present the Hamiltonian model, the basic
aspects of the numerical computations and we introduce the
thermodynamic and dynamical quantities that characterize the state of
the system. In Section \ref{SecAnaly} we compare the statistical
properties of our model to the usual ones, using both analytical and
numerical approaches. We derive the thermodynamical properties using
the hypothesis that the dynamics is dominated by random phased waves.
We then propose, in Section \ref{Sec:PT}, an Ising-like model based
on the observation that, above the transition, synchronized regions
of spins appear. The relation of the XY with an Ising-like model
allows us to describe the high temperature phase and to derive an
approximate value of the critical temperature. A brief summary of the
main results and conclusions is presented in Section \ref{SecConcl}.

\section{The Hamiltonian model, basic properties and numerical
computations}
\label{SecModel}

The XY model was introduced in statistical mechanics as a
two-dimensional version of the Heisenberg Hamiltonian. The spins are
fixed on the sites of a square lattice, and are characterized by a
rotation angle $\theta_{i}\in[-\pi,\pi]$
\begin{equation}
H_{XY}=J\sum_{(i,j)}^{N}[1-\cos(\theta_{i}-\theta_{j})]\;,
\label{Hxy}
\end{equation}
where $J$ is the coupling constant (with $J>0$ corresponding to the
ferromagnetic case, that we study here), $i$ and $j$ label the $N$
sites of a square lattice of side $\sqrt{N}$, i.e. $i=(i_x,i_y)$ with
$1\leq i_x,i_y \leq \sqrt{N}$ of coordinates $(x,y)$. The summation
is extended over all $i$ and its neighboring sites $j$. In the
following, without loss of generality, we set $J=1$ and the lattice
step to unity.

The spins evolve in time, $\theta_i=\theta_i(t)$, after 
adding a kinetic energy term to the XY Hamiltonian,
\begin{equation}
H=\sum_{i=1}^N\frac{p_i^2}{2}+H_{XY} \, ,
\label{Ho}
\end{equation}
where $p_i=\dot{\theta_{i}}$ is the spin momentum. The choice $J=1$
is equivalent to set time units and to rescale accordingly momentum
(in these units the ``inertia'' is also unity). With this kinetic
energy term the spins become in fact rotators, and the XY model
becomes a system of coupled rotators. The equations of motion are
\begin{equation}
\ddot{\theta_{i}}(t)=
-\sum_{j(i)}^4\sin[\theta_{i}(t)-\theta_{j}(t)] \;,
\label{dyn0}
\end{equation}
where the summation is over the four neighbors $j$ of site $i$. In
addition to the energy $H=E$, there exists a second constant of the
motion, the total angular momentum $P = \sum_i p_i$, which can be
chosen to be zero. We choose periodic boundary conditions in both the $x$
and the $y$ directions. Numerical integration of Eq.~(\ref{dyn0}) is
performed using the Verlet algorithm, which conserves the energy
${\cal O} (\Delta t^2 )$, $\Delta t$ being the time step, but exactly
preserves momentum and the symplectic structure.

Thermodynamical quantities are computed by averaging over time and
over the sites a single orbit (the evolution of the system from a
given initial condition). Typically, the system is started with a
Gaussian distribution of momenta and with all the spins pointing in
the same direction $\theta_i=\theta_0$, for reasons that are
clarified in the following. No strong dependence on the chosen
initial condition in this class was observed, but one could improve
statistically our results by averaging over many orbits with
different initial conditions and the same energy.

The temperature is computed through the average squared momentum
per spin
\begin{equation}
T=\frac{1}{N}\sum_{i=1}^{N} \overline{p_i(t)^2} \, ,
\label{temp}
\end{equation}
where the overbar stays for temporal averaging.

The thermodynamical state can be characterized by several macroscopic
variables: the internal energy per spin $h=h(T)=E(T,N)/N$ ($E$ being
the constant total energy of the system); the magnetization
${\bf M}={\bf M}(T,N)$, which, as mentioned in the Introduction,
vanishes for $N \to \infty$, but is sizeable for any finite $N$;
the density of topological defects $\rho_v$, or vortices, which
is intimately related to the mechanism underlying the phase
transition.

The magnetization ${\bf M}=(M_x,M_y)$ is given by
\begin{equation}
{\bf M} \equiv \overline{{\bf M}(t)}=
\frac{1}{N}\sum_{i=1}^{N}
\left(
\overline{\cos\theta_i}, \overline{\sin\theta_i}
\right) \, ,
\label{mag}
\end{equation}
and describe the mean orientation of the spin field. A topological
defect is identified, as usual, by computing the total angle
circulation on a given plaquette (the sum of the four plaquette
angles $\bmod[-\pi,\,\pi]$); when it equals $ \pm 2 \pi $ this
quantity identifies a positive (negative) unitary vortex on the
plaquette. The total density of vortices, or vorticity, which is an
intensive quantity, is then given by
\begin{equation}
\rho_v=\frac{1}{N}\sum_{[i,j]}(\theta_{j+1}-\theta_j) \bmod[-\pi,\,\pi]\,,
\end{equation}
where $[i,j]$ denotes the sites $j$ of the $i$-th plaquette. We have
studied the temporal evolution and the spatial distribution of the
vortices as a function of the temperature. Because of the periodic
boundary conditions and since $P=0$, the number of positive vortices
equals the number of negative ones.

\begin{figure}[htb]
\centering\epsfig{figure=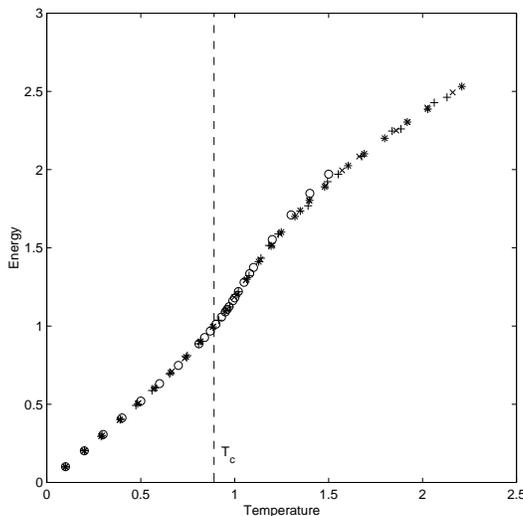,width=8cm}
\begin{minipage}[b]{15cm}
\caption{\baselineskip=12pt
Total energy per spin as a function of the temperature. At
low temperature the energy grows as $T$ and at high temperature the
energy tends to $T/2+2$. The Kosterlitz-Thouless phase transition
occurs at $T_{KT}\approx0.89$. $+$, {\sf x}, and * signs refer to the
Hamiltonian dynamical simulations for different lattice sizes,
$N=64^2$, $N=128^2$, and $N=256^2$ respectively; circles refer to
canonical Monte-Carlo simulations for a lattice size of $N=100^2$.}
\label{Fig1-h}
\end{minipage}
\end{figure}

We have performed simulations using various sets of parameters, to
study the system behavior depending on the temperature, the number of
spins, and the total time to test the stationarity of the relaxed
state. Having chosen parallel spins, one has an initially vanishing
potential energy, allowing the exploration of the low temperature
region through the reduction of kinetic energy. Random angles in the
$[-\pi, \pi]$ interval would give an energy per spin of 2, which
would then remain fixed producing a high temperature configuration,
as can be also deduced from the $h(T)$ function we compute below (see
Fig.~\ref{Fig1-h}). Therefore, initially, the magnetization is $|{\bf
M}|=1$.  Other initial conditions, with $|{\bf M}|=0$ were
investigated, for instance with half the spins oriented in the $x$
direction, and the other half in the $-x$ direction. After a
transient, the system relaxes to the same thermodynamical state.

In Fig.~\ref{Fig1-h}, we plot the total energy per spin $h(T)$. The
low ($T \ll 1$, $T\approx1$ corresponds to the value for which the
kinetic energy is of the same order of the potential energy) and high
($T \gg 1$) temperature behaviors are easy to understand.  At low
temperature equipartition of kinetic and potential energies gives
$h(T) \approx T$ (this is the linear regime, where angles between
neighboring spins are small). At high temperature, angles are
uniformly distributed in $[-\pi,\pi]$, the cosine interaction in
(\ref{Ho}) is negligible with respect to the kinetic energy, then
$h(T) \approx T/2+2$. These simple arguments suggest that for $T\ll
1$, the spin field is almost linear and can be represented as a
superposition of waves, which correspond to phonons, while for $T\gg
1$, the potential energy is negligeable, and this field becomes a set
of almost free fast rotators. The change in the behavior of the
$h(T)$ curve starts around the Kosterlitz-Thouless critical
temperature $T_{KT}$. The peak in the specific heat, related to the
second derivative of the energy, occurs instead at a somewhat higher
temperature, a fact well documented in Monte-Carlo computations.

\begin{figure}[htb]
\centering\epsfig{figure=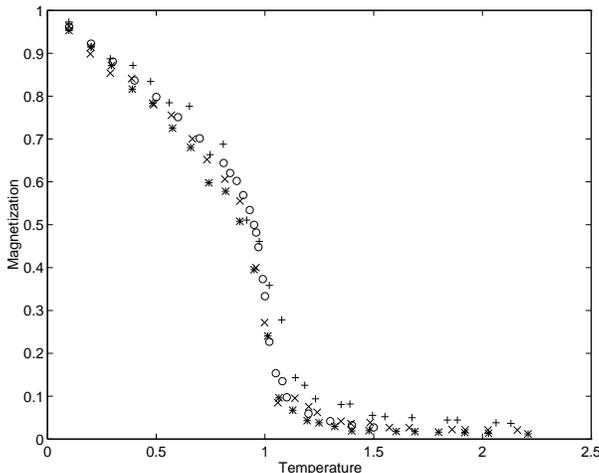,width=8cm}
\begin{minipage}[b]{15cm}
\caption{\baselineskip=12pt
Absolute value of the magnetization as a function of the
temperature. Finite size effects give a finite magnetization at low
temperatures. At low temperatures the magnetization decreases only
logarithmically with the number of spins. $+$, {\sf x}, and * signs
refer to the Hamiltonian dynamical simulations for different lattice
sizes, $N=64^2$, $N=128^2$, and $N=256^2$ respectively; circles refer
to Monte-Carlo simulations for a lattice size of $N=100^2$.}
\label{Fig2-M}
\end{minipage}
\end{figure}

Figure \ref{Fig2-M}, shows the absolute value of the magnetization as
measured using formula (\ref{mag}). Although in the thermodynamic
limit the magnetization must vanish, in a finite system and for low
temperature we expect an observable macroscopic magnetization, which
decreases logarithmically with the number of spins as noted by
Berezinskii \cite{Ber71}. This slow decrease with system size is
hardly observable in Fig.~\ref{Fig2-M}, but points corresponding to
larger sizes give systematically a smaller magnetization. In the high
temperature region, since angles are randomly distributed in space at
any time, the magnetization vanishes algebraically with the number of
spins.

\begin{figure}[htb]
\centering\epsfig{figure=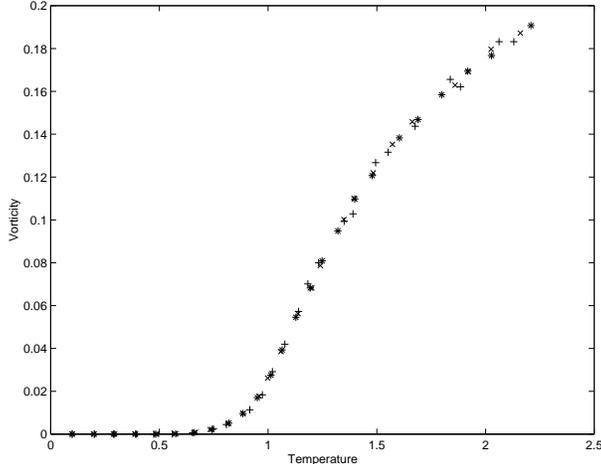,width=8cm}
\begin{minipage}[b]{15cm}
\caption{\baselineskip=12pt
Number density of vortices as a function of the temperature
normalized to the total number of spins.  This quantity is intensive.
$+$, {\sf x}, and * signs refer to the Hamiltonian dynamical
simulations for different lattice sizes, $N=64^2$, $N=128^2$, and
$N=256^2$ respectively.}
\label{Fig3-NVor}
\end{minipage}
\end{figure}

The behavior of vorticity with temperature, plotted in
Fig.~\ref{Fig3-NVor}, is related to the form of the $h(T)$ curve.
Indeed, at low temperature aligned angle configurations are typical
and consequently vorticity vanishes, $\rho_v\rightarrow0$.  In the
high temperature regime, spins are randomly distributed and the
vortex density approaches the asymptotic value
$\rho_v\rightarrow1/3$. A dramatic growth of the vorticity occurs in
the phase transition region (the temperature in the range $T\approx
0.8$ to $1.5$).

All our results are in perfect agreement with those of Monte-Carlo
simulations (see e.g. \cite{Bra93}), some of them are plotted in the
Figs.~\ref{Fig1-h} and~\ref{Fig2-M} for comparison. Therefore,
microcanonical (Hamiltonian dynamics) and canonical (Monte-Carlo)
computations give, at least for the class of initial conditions
studied here, the same thermodynamical equilibrium states.

\section{Dynamical and statistical properties}
\label{SecAnaly}

We know that at low temperatures, the main contribution to the
canonical partition function comes from the configurations in which
the spins are almost aligned, configurations where the angle
differences $\theta_i-\theta_j$ are small.  In such a situation, the
equations of motion (\ref{dyn0}) can be linearized, and the spin
field is therefore represented as a superposition of linear waves. We
will here determine the effective dispersion relation for these waves
at low temperature using consistency relations with temperature and
internal energy. Let us introduce a representation of the spin angles
in the form of a random phased field
\begin{equation}
\theta_i=
    \sum_k \alpha_k\cos(\psi_k^{i})\,,\;
\psi_k^{i}=
    k \cdot x_i-\omega_k t+\phi_k\;,
\label{rand}
\end{equation}
where the summation is over the wave-vectors $k= (k_x,k_y)=
2\pi(n_x,n_y)/\sqrt{N}$, with $n_x,n_y=1,\ldots,\sqrt{N}$ integers;
the wave spectrum is given by the $\alpha_k$, and the phases $\phi_k$
are supposed to be random, uniformly distributed in the circle. We
also should have added to (\ref{rand}) a term of the form $\Omega_i
t$ reflecting the individual rotation of spins with some frequency
$\Omega_i$, which in fact can be considered a function of the
temperature. However, at low temperature most of these frequencies
must vanish, because the energy needed to trigger the free rotation
is only reached at energies of the order of $h=2$ (or temperatures of
the order of $T=1.5$, as can be seen in Fig.~\ref{Fig1-h}), when a
spin, considered as a perturbed pendulum, crosses the separatrix.
Then, in the following we put $\Omega_i=0$ in the low temperature
(this term becomes important at high temperatures).

Although the main assumption in (\ref{rand}), that is the random
character of the wave phases, may only be justified {\em a
posteriori}, we performed some numerical tests which showed that this
ansatz is consistent with the properties of the system (at low and
also at finite temperatures). We found that the probability
distribution of the spin velocities is almost uniform, and that
moreover, the motion of one single spin is ergodic (in the sense that
its temporal mean and variance coincide with the spatial ones). In
addition, if the system were linear this assumption is equivalent to
the assumption of thermodynamic equilibrium (with a bath at
temperature $T$). In fact this is the case at very low and very high
$T$; of course, in the intermediate range the influence of the
nolinearity becomes important, for instance, establishing non-local
interactions or some type of self-organization, in which case this
assumption would not be necessarly valid.

In formula (\ref{rand}) there are two unknown functions of the
wave-number which remain to be determined, the wave frequencies
$\omega_k$ and the spectrum $\alpha_k$. Note, in addition, that
(\ref{rand}) is in fact a change of variables of the angles
$\theta_i$ to the amplitudes $\alpha_k$, where implicitly, a slow
temporal dependence may be included, the fast time dependence being
assured by the phase $\omega_kt$. In this sense, formula (\ref{rand})
is rather general and can take into account a large variety of field
states.

The definition of the temperature imposes a constraint on the form of
the wave spectrum. Indeed, replacing formula (\ref{rand}) into the
definition of the temperature (\ref{temp}) we obtain
\begin{eqnarray}
T&=&\frac{1}{N}\sum_i^N\langle{\dot\theta}_i^2\rangle\nonumber\\
 &=&\frac{1}{2}\sum_k\alpha_k^{2}\omega_k^{2}\;,
\label{jeansT}
\end{eqnarray}
where $\langle\cdots\rangle$ stays for averaging over the random
phases $\phi_k$, and where we used the identity
$$
\langle\cos(\psi_k^{i})\cos(\psi_{k'}^{i})\rangle=
   \frac{1}{2}\delta_{k,{k'}}\;,
$$
being $\delta_{k,{k'}}$ the Kronecker symbol. For instance, if
equipartition of the energy among different degrees of freedom is
assumed, one obtains the usual Jeans spectrum, given by
\begin{equation}
\alpha_k^{2}=\frac{2T}{N\omega_k^{2}}\;,
\label{jeans}
\end{equation}
which means that each degree of freedom takes a fraction $T/2$ of the
total kinetic energy. In fact, as we will show below, the detailed
knowledge of the spectrum turns out not to be necessary to derive the
dispersion relation and compute thermodynamical quantities.

Before proceeding to the computation of the dispersion relation, let
us derive a general expression of the energy per spin $h$,
\begin{equation}
h=\frac{T}{2}+2-\frac{1}{N}
  \sum_{<i,j>}^{N}\langle\cos(\theta_{i}-\theta_{j})\rangle\;,
\label{h}
\end{equation}
where the first term comes from the definition of the temperature,
the second term is the constant added for convenience to the
Hamiltonian to make the energy vanish at zero temperature, and the
last term includes, for each lattice site, a summation over two
neighbors, for instance west (in the $x$ direction) and south (in the
$y$ direction). When the expansion (\ref{rand}) is introduced into
Eq. (\ref{h}) the typical contribution to the sum in the third term
on the r.h.s. is
$$
U_N^{(i,j)}=\left<
    \cos\left[\sum_{n=1}^{N}\;
        A_{k_n}^{(i,j)}\sin(B_{k_n}{(i,j)}+\phi_{k_n})\right]
    \right>\;,
$$
where $n$ relabel the modes and
\begin{eqnarray}
A_{k_n}^{(i,j)}&=&
        2\sin\left[\frac{k\cdot (x_i -x_j)}{2}\right]\nonumber \\
B_{k_n}^{(i,j)}&=&
        \frac{k \cdot (x_i +x_j)}{2}+\omega_k t\nonumber
\end{eqnarray}
are local in $n$ and independent on the phases $\phi_{k_n}$.  Let us
split the summation over the wave-numbers into two terms, one
containing the phase, $\phi_{k_N}$, and the other with the remaining
$n=1,\ldots,N-1$ phases. Since
\begin{eqnarray}
\left<\cos\left[
   A_{k_N}^{(i,j)}\sin(B_{k_N}^{(i,j)}+\phi_{k_N})
\right]\right>&=&J_0(A_{k_N}^{(i,j)})\nonumber \\
\left<\sin\left[
   A_{k_N}^{(i,j)}\sin(B_{k_N}^{(i,j)}+\phi_{k_N})
\right]\right>&=&0\;,
\end{eqnarray}
where $J_0$ is the Bessel function of zero order, we obtain the
recursion relation
$$
U_N^{(i,j)}=J_0(A_{k_N}^{(i,j)})\, U_{N-1}^{(i,j)}\;,
$$
which readily gives the result
$$
U_N^{(i,j)}=\prod_n\, J_0(A_{k_n}^{(i,j)})\;.
$$
We finally get
\begin{equation}
h=\frac{T}{2}+2-\prod_k J_0(\beta_x)-\prod_k J_0(\beta_y)\;,
\label{hexpr}
\end{equation}
where $\beta_x=2\alpha_k\sin k_x/2$ and $\beta_y=2\alpha_k \sin
k_y/2$ and the last two products come from north/south and east/west
neighbours on the lattice. Eq.~(\ref{hexpr}) is the general
expression for the energy per spin, as long as the system is
dominated by random phased waves. We note that in the case of a
symmetric spectrum $\alpha_{k_x,k_y}=\alpha_{k_y,k_x}$ (this is the
case for the Jeans spectrum, when the dispersion relation satisfies
$\omega_{k_x,k_y}=\omega_{k_y,k_x}$), the products of zero order
Bessel functions are equal,
\begin{equation}
\frac{v}{2}=\prod_k J_0(\beta_x)=\prod_k J_0(\beta_y)\;,
\label{Eq-v}
\end{equation}
where $v$ is related to $h$ by $h(T)=T/2+2-v(T)$, and is the average
potential energy. In order to obtain an explicit form of the energy
per spin as a function of the temperature, $h=h(T)$, we must
determine in a self-consistent way the dispersion relation; this may
be done by using the equation of motion in its full nonlinear form.
But let us begin with the linear case, which will serve as a guide to
the self-consistent computation. The linear equation of motion reads
\begin{equation}
\ddot\theta_i=\sum_{<j>}\, \theta_i-\theta_j\;,
\end{equation}
Substituting (\ref{rand}), one obtains
\begin{eqnarray}
-\sum_k\,
   \omega_k^2\alpha_k\cos\psi_k^{i}&=&
   -\sum_k\,\beta_x\,
   [\sin(\psi_k^{i}+\frac{k_x}{2})-
    \sin(\psi_k^{i}-\frac{k_x}{2})]+\nonumber\\
   &&-\sum_k\, \beta_y
   [\sin(\psi_k^{i}+\frac{k_y}{2})-
    \sin(\psi_k^{i}-\frac{k_y}{2})]
\label{lin}\\
   &=&-4\sum_k\, \alpha_k\,
   [\sin^2(\frac{k_x}{2})+\sin^2(\frac{k_y}{2})]\,
   \cos \psi_{k}^{i}\;,             \nonumber
\end{eqnarray}
although the linearity of the equation allows an identification term
by term in the summation over $k$, a different strategy for the
solution would be to average (\ref{lin}) over all the phases but one,
e.g. over the phases $\phi_{k'}\not=\phi_k$ after multiplying both
sides by $\cos\psi_k^{i}$, in order to isolate only one term in the
summation on the left hand side. One finally obtains the linear
dispersion relation
\begin{equation}
\omega_k=\omega_{0k}=
   4(\sin^2\frac{k_x}{2}+\sin^2\frac{k_y}{2})\;,
\label{lindis}
\end{equation}
The frequency spectrum is symmetric with respect to the exchange $k_x
\leftrightarrow k_y$; in the following we assume, without loss of
generality, that the spectrum is symmetric with respect to this
transformation.

This procedure can be generalized to the nonlinear case, although the
r.h.s. of Eq.~(\ref{dyn0}) is no more local in $\phi_k$.  After
substituting the expansion (\ref{rand}) into Eq.\ (\ref{dyn0}), we
must average over the phases $\phi_{k'}\not=\phi_k$, which we denote
$<\cdots>'$, terms of the form
\begin{equation}
\left<\sin(\theta-\theta_E)\right>'= -
   \left<
   \sin\left[
   \sum_k\,\beta_x
   \sin(\psi_k+\frac{k_x}{2})
   \right]\right>'\:,
\label{east}
\end{equation}
where $\theta_E$ is e.g. the east neighbor of angle $\theta$, having
a phase $\psi_E=\psi_k+k_x=k \cdot x+k_x-\omega_{k} t+\phi_k$. We now
split the summation in (\ref{east}) into a term containing the phase
$\phi_k$ and the others, and develop the sine of the summation of two
terms into a product. The averaging of these two terms reduces to
\begin{equation}
\left<
   \cos\left[
   \sum_{k'}\,\beta_x
   \sin(\psi'_k\pm \frac{k'_x}{2})\right]
   \right>'=\frac{v}{2J_0(\beta_x)}\;,
\end{equation}
because the averaging of the sine term is zero. The two neighbors in
the $x$ direction (west and east sites) give the following expression
\begin{eqnarray}
\left<\sin(\theta-\theta_E)\right>'+&&
\left<\sin(\theta-\theta_W)\right>'=    \nonumber\\
   &&-\frac{v}{2J_0(\beta_x)}\left\{
   \sin\left[\beta_x\sin(\psi_k+\frac{k_x}{2})\right]-
   \sin\left[\beta_x\sin(\psi_k-\frac{k_x}{2})\right]
   \right\}\;.
\label{ew}
\end{eqnarray}
Those in the $y$ direction give exactly the same contribution with
$\beta_x \rightarrow \beta_y$. On the other hand, the left hand side
of Eq.~(\ref{dyn0}), the temporal part of the equation of motion, is
reduced, after substitution of the random waves expression, to
\begin{equation}
\left<\ddot\theta_i\right>=
   -\omega_k^2\alpha_k\cos\psi_k^i\:.
\label{ddt}
\end{equation}
As before, in the linear computation, we multiply both sides of the
equation of motion (\ref{ddt}) and (\ref{ew}) (including the similar
terms for the north and south neighbors), by $\cos(\psi_k^i)$ and
average over the phase $\phi_k$, noting that this average involves
first order Bessel functions $J_1$. We finally find the desired
dispersion relation
\begin{equation}
\omega_k^2 \alpha_k=
   2v\left[
   \sin (\frac{k_x}{2})\,\frac{J_1(\beta_x)}{J_0(\beta_x)}+
   \sin (\frac{k_y}{2})\,\frac{J_1(\beta_y)}{J_0(\beta_y)}
   \right]\;.
\label{nldr}
\end{equation}
This dispersion relation is nonlinear, i.e. the frequency depends on
the spectrum amplitudes, in two ways: implicitly through the
$\alpha_k$s in $v$ and explicitly in the Bessel functions. To go
further, let us investigate the dependence of the arguments of the
Bessel functions on the parameters $T$ and $N$, and the related
limiting form of the total energy per spin $h$.  We know that $v$,
being an intensive thermodynamic quantity, does not depend
on the number of spins (the size of the system); moreover, the linear
frequency (\ref{lindis}) is bounded from below, $\omega_{0k}^{2}\sim
\sin^2(k_x/2)>{\cal O}(1/N)$. We also note, as it is natural for a
system near an equilibrium state, that a large number of $k$ modes
must contribute to the energy of the system, and then in general we
have $\alpha_k^{2}\sin k_x/2\rightarrow 0$ when $N\rightarrow
\infty$, for a large range of temperatures. If this were not the
case, the energy would be concentrated in a few high $k$ modes, which
is clearly in contrast with the observations. For the Jeans spectrum,
one finds specifically $\alpha_k^{2}\sin k_x/2\approx{\cal
O}(T/\sqrt{N})$. Using these approximations we can now develop the
logarithm of the product of the Bessel functions in (\ref{Eq-v}) to
obtain
\begin{equation}
v=2\exp\left\{-\frac{1}{8}
\sum_k\alpha_k^{2}\omega_{0k}^{2}
\right\}\;.
\label{v}
\end{equation}
In the same approximation, for fixed temperatures and large lattices,
the dispersion relation (\ref{nldr}) reduces to
\begin{equation}
\omega_k^{2}=\frac{v\omega_{0k}^{2}}{2}\;.
\label{nldr1}
\end{equation}
Introducing this expression into the formula (\ref{v}) for $v$ and
using the definition of the temperature in terms of random phased
waves (\ref{jeansT}) we obtain an implicit equation for the potential
energy,
\begin{equation}
v=2\exp\left[-\frac{T}{2v}\right]\;.
\label{vim}
\end{equation}
We notice that the nonlinear dispersion relation reduces to the
linear one for $T=0$. The nonlinear effects, taken into account in
(\ref{nldr1}), appear as a renormalization of the phonon frequency
(energy) due to the coupling of the phonons with a thermal bath
created by the other phonons like in a mean field. We also note that
the factor 2 found in the right hand side of (\ref{vim}) and in the
exponent can be related to the lattice dimensionality (it results
from the addition of the $x$ and $y$ terms). Moreover, as we
anticipated, formula (\ref{vim}) does not depend explicitly on the
form of the spectrum $\alpha_k$.

\begin{figure}[htb]
\centering\epsfig{figure=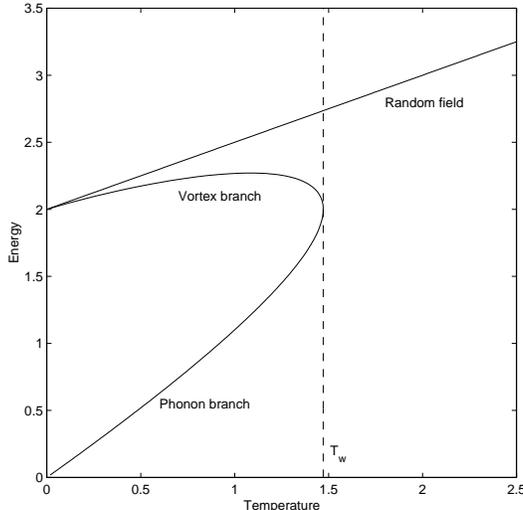,width=8cm}
\begin{minipage}[b]{15cm}
\caption{\baselineskip=12pt
Plot of the energy density versus temperature computed from
Eq.~(\ref{vim}). The straight line characterize the random field
state, the lower branch, the phonons (spin waves), and the upper
branch, lattice excitations of the vortex dipole type.}
\label{hth}
\end{minipage}
\end{figure}

Let us now investigate the implicit equation (\ref{vim}) by solving
it numerically. In Fig.~\ref{hth}, where the energy density $h$
versus the temperature $T$ is plotted using formulas $h=T/2+2-v(T)$
and (\ref{vim}), we see three branches according to the solutions of
(\ref{vim}). Although the three branches are in principle acceptable
solutions, they would have different ``weights''. For instance, if
one associates to each branch a Gibbsian probability, proportional to
$\exp(-h_i/T)$, where $h_i$ denotes the branch energy $i=0, 1, 2$,
the lower branch has, at a given temperature, the higher probability.
The choice of a Gibbsian probablity distribution of states is not in
contradiction with our microcanonical approach. Indeed, the energies
$h_i$ are related to {\em macroscopic} equilibrium states, and the
Gibbs probabilities represent the best choice, taking into account
the usual constraints of maximal entropy and normalization.

The straight line corresponds to the solution $v=0$, which represents
a random spin field. Indeed, in the high temperature region, since
the potential energy is bounded, we can expect, that the rotators are
freely rotating without any order, {\em i.e.} that
$(\theta_{i}-\theta_{j})$ is random with constant distribution in
$[0, 2\pi)$. This gives us an energy density $h=T/2+2$ and $v(T)=0$.
The implicit solution of (\ref{vim}) extrapolates thus the good
asymptotic behavior of the energy for high temperatures to the low
temperature region. This solution exists for any temperature, but in
the low temperature region spin configurations associated to this
branch should be unstable (in the sense that for general initial
conditions, as in our numerical computations, the system cannot
evolve to this state).

\begin{figure}[htb]
\centering\epsfig{figure=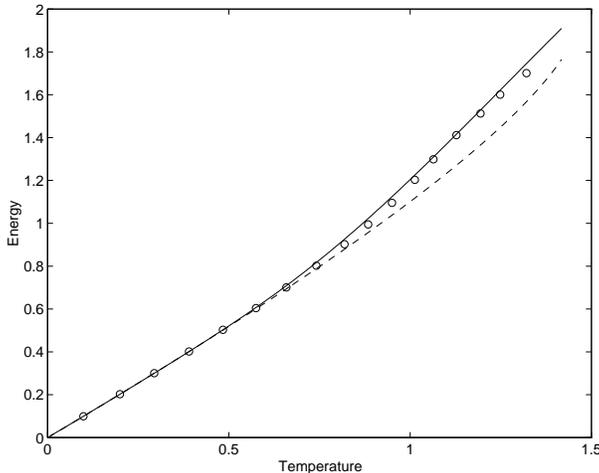,width=8cm}
\begin{minipage}[b]{15cm}
\caption{\baselineskip=12pt
Plot of the energy density versus the temperature. The
circles refer to the numerical data, the dashed line to the phonon
branch of Eq.~(\ref{vim}), and the solid line to the theoretical
averaged density of energy, Eq.~(\ref{hstatis}).}
\label{hthbis}
\end{minipage}
\end{figure}

Below a certain temperature $T=T_{w}$ we also see that two other
branches appear. As the lower branch has the higher probability, in
the low temperature region, we may expect the lower energy branch,
which we call the phonon branch, to be the physical relevant
solution, and the third branch to have no physical meaning at all.
However, if we now consider that in fact the lattice is populated
with two types of species of different energies $h_1(T)$ and $h_2(T)$
corresponding to the lower (phonons) and upper (which we call vortex)
branch respectively, we obtain
\begin{equation}
h=\frac{h_1 e^{-2h_1/T}+h_2 e^{-2h_2/T}}
       {e^{-2h_1/T}+e^{-2h_2/T}}  \:,
\label{hstatis}
\end{equation}
which is almost in total agreement with the numerical data up to
$T_{w}$, as we can see in Fig.~\ref{hthbis}.

The upper branch can be related to ``dipole vortices'' since it
starts at an energy equal to $2$ (see Fig.~\ref{hth}) corresponding
to a lattice composed by topological defects forming a perfect
crystal ($\theta_i-\theta_j=\pm\pi/2$ and $\sum_{j(i)}\sin(\theta_i -
\theta_j)=0$ overall the lattice). This crystal state, which is in
fact a periodic array of dipoles, is a stationary solution of the
equations of motion at $T=0$, around which no linear development can
be made (the linear term is zero, vortices can be considered as
nonlinear particles). The factor $2$ in the exponents of
(\ref{hstatis}) takes into account that vortices come in pairs
(dipoles at low temperature) on the lattice in order to conserve the
total circulation.

The temperature $T_{w}$ determines the upper border of the wave
dominated regime, above this temperature the wave branch disappears
(as well as the vortex dipole branch). In order to determine
precisely $T_{w}$ it is useful to rewrite (\ref{vim}), as a formula
for the temperature in terms of $v$,
\begin{equation}
T=-2v\log\left(v/2\right)\: .
\end{equation}
If we now calculate the extremum of $T(v)$ for $0<v<2$, we obtain
for $v=2/e=T_w/2$, $T=T_{w}=4/e\approx 1.47$, which is exactly
the critical temperature found in \cite{Stu82} using a Hartree-Fock
approximation. This temperature also appears with the use of
renormalization techniques in \cite{Spi93}. In this case, we note
that the equation for $v$ is formally identical to the one for the
renormalization constant in this renormalization group calculation.
However, the two methods are basically different, and in particular
the relation between $v$ and the dispersion relation (\ref{nldr1}),
allows us to get more precise informations on the different
thermodynamical quantities of the system. For instance, as mentioned
above, the relation between $v$ and $h$, which is different from the
relation between the renormalization constant and the energy, leads
to a very good description of the $h(T)$ curve up to $T_w$.

\begin{figure}[htb]
\centering\epsfig{figure=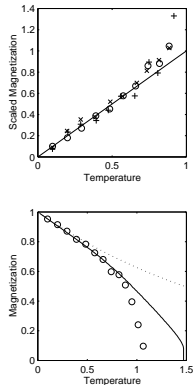,width=6cm}
\begin{minipage}[b]{15cm}
\caption{\baselineskip=12pt
Finite size effects on the magnetization. The upper plot shows $-4\pi
v\log(M)/\log(2N)$ as a function of the temperature; the line refers
to formula (\ref{maglt}); $+$, {\sf x}, and {\sf o} refer to
numerical data for lattice sizes of $N=64^2$, $N=128^2$, and
$N=256^2$, respectively. The lower plot shows $M(T)$,  circles are
for a $N=256^2$ lattice; the thin line is the analytical result and
the dotted line is the result using only the contribution of linear
spin waves: $M=\exp(-T\log(2N)/8\pi)$.}
\label{magth}
\end{minipage}
\end{figure}

Let us now evaluate the magnetization for a finite size system with
periodic boundaries using the same approach as before: we assume that
the spin field can be represented by a superposition of random phased
waves (\ref{rand}). We call $\theta_0$, the average over the lattice
of the $\theta_i$, which is a constant of motion, since the total
momentum is conserved. An average of the magnetization over the
random phases brings up to a similar calculation as the one for the
energy per spin and leads to,
\begin{equation}
\left<{\bf M}\right> =\prod_kJ_0\left(\alpha_k\right)
   \left(\cos\theta_0,\sin\theta_0\right)
\end{equation}

This expression remains exact as long as the random phases
approximation is valid. In order to develop the logarithm of this
expression, we have to take into account that $\alpha_{k}^{2}<{\cal
O}(T)$, which implies that the development is only valid in the low
temperature regimes $T\ll 1$. Moreover a detailed knowledge of the
spectrum is also required. Therefore, considering the observed almost
flat spectrum of the momentum, we assume the equipartition of the
kinetic energy among the modes and use the Jeans spectrum given in
(\ref{jeans}). The absolute value of the magnetization, $\langle|{\bf
M}|\rangle$, is given by the product over the different $k$ of the
Bessel functions. In the low temperature regime using (\ref{nldr1})
and (\ref{jeans}), we obtain for its logarithm,
\begin{equation}
\log\left(\langle|{\bf M}|\rangle\right)=
   -\sum_k \frac{\alpha_k^2}{4}=
   -\frac{T}{2N}\sum_k \frac{1}{\omega_k^2}=
   -\frac{T}{v}G\left(0\right)\;,
\label{logm}
\end{equation}
where $G$ is the Green function of the linear wave equation, $G(r)=
\sum_k\exp(ik\cdot r)/\omega_{0k}^2$, with $G(0)=(1/4\pi)\log(2N)$.
The expression for the magnetization is then
\begin{equation}
\left<{\bf M}\right>=\left(\frac{1}{2N}\right)^{\frac{T}{4\pi v}}
\left(\cos\theta_0,\sin\theta_0\right)\;.
\label{maglt}
\end{equation}

A plot of this expression, scaled in order to obtain a function of
the temperature, $-4\pi v\log(M)/\log(2N)$, is shown in
Fig.~\ref{magth} (top). Within the errors of the numerical data, the
points collapse, at low temperature, to a unique curve. The result
(\ref{maglt}), obtained including the correction due to the nonlinear
contribution to the dispersion relation, substantially improves the
usual estimation based on the linear wave approximation (see the
bottom of the figure). The agreement of the theoretical results and
the numerical data can be made more precise by taking into account
the different energy branches. Indeed, if we now take into account
that the lattice is populated by two types of species, we can
consider that only the phonons contribute to the magnetization,
vortex dipoles having total magnetization zero. As the density of
phonons is given by $n_1=e^{-2h_1/T}/(e^{-2h_1/T}+e^{-2h_2/T})$, the
observed magnetization must then be simply $n_1 {\bf M}$. The
absolute value of this quantity is plotted in Fig.~\ref{magth}
(bottom). The agreement with numerical data is valid up to the
critical temperature.

\begin{figure}[htb]
\centering\epsfig{figure=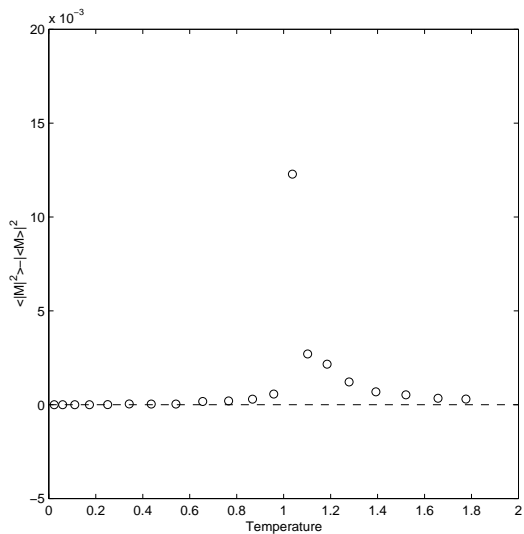,width=8cm}
\begin{minipage}[b]{15cm}
\caption{\baselineskip=12pt
Plot of the variance of the magnetization versus
the temperature ($N=128^2$). The transition region ($T\approx 1$) is
clearly identified.}
\label{varm}
\end{minipage}
\end{figure}

In the same way we compute the averaged absolute value of the
magnetization
$$
\langle | {\bf M }|^2\rangle=
\frac{1}{N^2}\sum_{i,j}\langle\cos(\theta_{i}-\theta_{j})\rangle\;,
$$
and we obtain the expression
\begin{equation}
\langle | {\bf M }|^2\rangle=\frac{1}{N^2}\sum_{i,j}
\prod_kJ_0\left(2\alpha_k \sin\left({
k}\cdot\frac{{x}_i-{x}_j}{2}\right)\right)\;,
\label{varmag}
\end{equation}
which, in the low temperature regime, assuming a Jeans spectrum, and
making the same expansion as in (\ref{logm}), leads to
$$
\langle|{\bf M }|^2\rangle=
   \frac{1}{N^2}\sum_{i,j}
   \exp\left(-\frac{2T}{v}G\left( r_{ij}\right)\right)\;,
$$
where $r_{ij}=x_i-x_j$. A first order calculation leads to the same
result as (\ref{maglt}), implying that the variance $\mbox{Var}({\bf
M})\equiv\langle |{\bf M }|^2\rangle-|\langle{\bf M}\rangle|^2$
vanishes at leading order in the temperature. However, at finite
temperature, the two values can differ, and the variance of the
magnetization can have a nonzero value. This variance can moreover be
used to characterize the phase transition, since it is the average
over the lattice of the correlation function $\langle
e^{\theta_{i}-\theta_{j}}\rangle$; it is also related to the
susceptibility $\chi\equiv(N/T)\mbox{Var}({\bf M})$. Indeed, in
Fig.~\ref{varm} we plot the magnetization variance as a function of
the temperature, and find that up to temperatures of the order of the
Kosterlitz-Thouless $T_{KT}$, the variance remains very small, and
suddenly it grows around $T_{KT}$. This behavior suggests that the
magnetization is distributed almost like a delta function, in the low
temperatures regime (all the spins are pointing to the same
direction), and that near the transition, the amount of randomly
oriented spins increases dramatically. Higher order effects on the
variance of the magnetization can be computed using the identity
$$
J_0(2z\sin\alpha)=J_0^2(z)+
   \sum_{m=1}^{\infty}J_m^2(z)\cos(2m\alpha)\;,
$$
which gives, after replacing it into (\ref{varmag}),
$$
\mbox{Var}({\bf M})=
   |\langle{\bf M}\rangle|^2\left(
   \frac{1}{N^2}\sum_{i,j}
      \prod_k\left[1+M_k(r_{ij})\right]-1\right)\;,
$$
where
$$
M_k(r_{ij})=\sum_m\frac{J_m^2(\alpha_k)}{J_0^2(\alpha_k)}
\cos\left(m k\cdot r_{ij}\right)\;.
$$
To the first nonvanishing order ($m=1$) we obtain
$$
\mbox{Var}({\bf M})=
   2\left(\frac{T}{vN}\right)^2|\langle{\bf M}\rangle|^2
   \sum_k\frac{1}{\omega_{0k}^4}\;.
$$
This allows us to compute an approximate value of the suceptibility
$\chi$,
$$
\chi\approx 3.87\,10^{-3}  
      \frac{T}{v^2}\left( 2N \right)^{(1-T/2\pi v)}\;.
$$
A similar result can be found in \cite{Arc97}, in the limit
$v\rightarrow 2$.

Concerning the density of vortices, we can see in
Fig.~\ref{Fig3-NVor} that around $T\approx 1.3$ the curve has an
inflexion point. This change in the number of defects may be
explained as follows. Let us consider a single plaquette, a defect
may appear when the four angles are in increasing (decreasing) order,
and the last angle is larger (smaller) than $\pi$ ($-\pi$), if the
first angle is set to zero. In our model, due to the continuous
symmetry group, no particular direction is favored, and therefore the
successive differences between the angles $\theta_i-\theta_j$
(discrete gradients) are all equivalent. A defect is then obtained
when these gradients satisfy (in average)
$\langle|\theta_i-\theta_j|\rangle>\pi/3$. At low temperatures the
amplitudes of the local gradients are determined by the phonons and
become steeper at higher temperatures, reaching a point where they
are large enough to generate the vortices. Using the previous
results, we calculate the temperature for which this condition is
reached,
\begin{equation}
\langle\left(\theta_i-\theta_j\right)^2\rangle=
  2\sum_k\alpha_k^2 \sin^2\frac{k_x}{2}=
      \frac{T}{v(T)}=
      \frac{\pi^2}{9}\:.
\label{tcvort}
\end{equation}
If we substitute now this result into equation (\ref{vim}), we find
\begin{equation}
T=\frac{2\pi^2}{9}\exp\left({-\frac{\pi^2}{18}}\right)
\approx 1.27\:,
\end{equation}
which is in very good agreement with the position of the inflexion
point in the numerical data. One may consider that around this
temperature a proliferation of vortices should occur.

\section{The phase transition}
\label{Sec:PT}

The usual physical picture of the Kosterlitz-Thouless transition is
based on the unbinding of vortex pairs: below the critical
temperature long range correlations are established by spin waves,
the phonons in our Hamiltonian model; above the critical temperature
long range order is destroyed by the proliferation of free vortices.
This simple physical picture of the mechanism of the transition does
not exhaust the complex processes related to the appearence of a kind
of self-organization in the system: the vortex distribution is not
uniform, vortices form clusters which separate domains of relatively
well ordered spins. In this Section we study, using numerical
simulations, the spatial distribution of topological defects and the
equilibrium configurations of the system, using an Ising-like model,
in order to describe the partial order which is still present above
$T_{KT}$. 

In the previous Section we showed that the ordered state of the
system is dominated by phonons, and that a second type of
excitations, the vortex dipoles, is also present. The energies of the
two branches meet at a temperature $T_w$, above which both modes
disappear, suggesting that a phase transition to a disordered state,
might occur. Within this picture, $T_w$ would be thus the temperature
where the unbinding of vortex dipoles appears, destroying the long
range order created by the waves. However, the actual transition
temperature $T_{KT}$ is much lower than $T_w$, indicating that other
processes, not included in the representation (\ref{rand}) of the
spin motion, take place. 

Indeed, in this low temperature analysis we have omitted the rotation
of the spin, excluding terms of the form $\Omega_i t$. Moreover, most
significantly, we have assumed that the phases of the spin field were
random and uncorrelated from site to site, thus neglecting the
influence of organized spin motion. Although the proliferation of
vortices breaks the long range order, the spin field can still be
organized into domains (where the spins rotate synchronously)
separated by lines of vortices. The phase transition would be
associated in this case, with the appearence of a self-organized
state,
where, although long range order is absent, a kind of partial local
order, established by separated domains of coherent spins, sets in.

\begin{figure}[htb] \centering\epsfig{figure=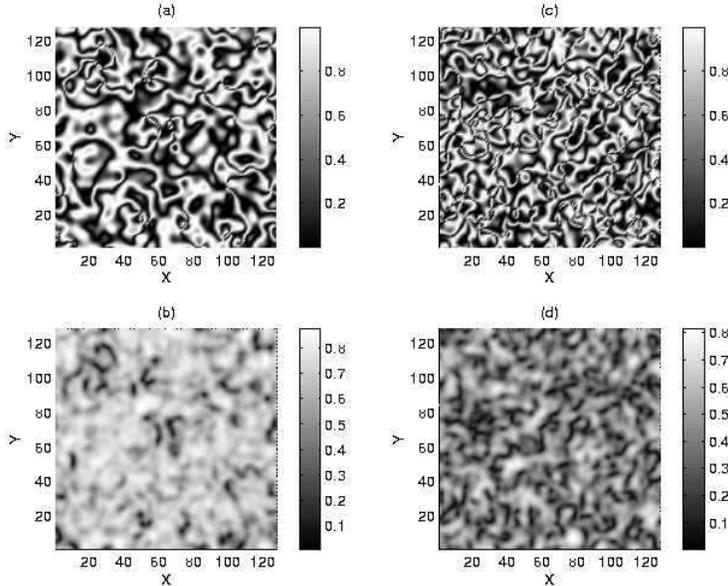,width=10cm} \begin{minipage}[b]{15cm} 
\caption{\baselineskip=12pt Spatial distribution of the local magnetization, showing the domains and the interfaces (see text). (a) and (c) orientation ($\sin^2(\theta_i^l/2)$), one passes from white to black by a rotation of $\pi/4$; (b) and (d) intensity ($|{\bf m}_i^{(l)}|$), black are disordered regions. (a) and (b): $T=1.02$, (c) and (d): $T=1.29$.}
\label{vortpos}
\end{minipage}
\end{figure}

We consider this possibility and study the spatial distribution of
defects around the critical temperature. By direct visualization of
the spin field ${\bf m}_i=(\cos\theta_i,\sin\theta_i)$, we noted that
the number of isolated defects is negligible, and that they often
appear to be in small clusters along the domain borders (this may
already be seen in Fig.~7 of Ref.~\cite{Tob79}). The problem one
encounters with direct visualization is that the spins move too fast
to
allow the observation of coherent structures. To gain some further
insight we introduce then a new diagnostic to correctly visualize the
spatial structures. We update each spin of the lattice with one fifth
of the sum of itself and its four nearest neighbors and repeat this
operation a few times, to obtain an effective local magnetization
centered on the considered spin. This local average magnetization is
then defined by the iteration:
$$
{\bf m}_i^{(n+1)}=\frac{1}{5}
    \left[
        {\bf m}_i^{(n)}+\sum_{j(i)}^4 {\bf m}_j^{(n)}
        \right],
        n=1,\ldots,l\;,
$$
where $j(i)$ are the four neighbours of site $i$, and the number
of iterations is typically $l=10$. The resulting field ${\bf
m}_i^{(l)}$ is smoother than the initial field ${\bf m}_i^{(1)}$; at
these temperatures the amplitude of spin motion is large, and it
defines a direction at a given site, weighted by the orientation of
the surrounding spins. 

The effective spin field is shown in Fig.~\ref{vortpos}. The upper
plots are linear grey scale images of the quantity
$\sin^2\left(2\theta_i^{(l)}\right)$ (see also~\cite{Yur93}), where
$\theta_i^{(l)}$ is the angle of the local magnetization, and the
bottom ones are images of the quantity $|{\bf m}_i^{(l)}|$. In the
upper plots we can easily locate the vortices by looking at pinching
of the darker and brighter areas. They appear to be all bounded in
dipoles or chains of dipoles. The presence of these chains can be
interpreted as the birth of interfaces separating domains with local
order. These latter are more visible in the bottom images, where the
brighter regions characterize strong local magnetization, and
therefore locally aligned spins; while the darker regions, where the
orientation of spins change rapidly with position, indicate the
presence of vortex defects and disorder. In the left images,
representing a system of $N=128^2$ spins at a temperature slightly
above the critical temperature, $T=1.02$, we see that the disordered
regions are highly concentrated and that they tend to connect
themselves along lines. This tendency is even more clear in the right
images, where we show a system at $T=1.29$. We also note that the
size of the ordered regions, the domains, are much smaller at higher
temperatures. At $T=1.02$ the domains occupy connected regions of a
size comparable to that of the whole system, at higher temperature
the domains are instead confined in localized regions.

\begin{figure}[htb]
\centering\epsfig{figure=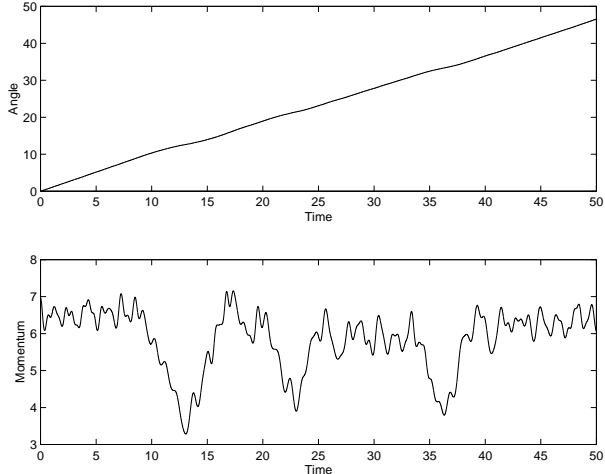,width=8cm}
\begin{minipage}[b]{15cm}
\caption{\baselineskip=12pt
Typical temporal behavior of a rotator at high temperature.
Top, spin $\theta_i(t)$ at the central site for $T=4.100$,
measured in units of $2\pi$. The spin rotates in the same
direction and with an almost
steady speed for long periods of time.
Bottom, plot of $\dot\theta_i(t)$ showing that its
fluctuations are fast
compared to the evolution of $\theta_i$.}
\label{timet}
\end{minipage}
\end{figure}

\begin{figure}[htb]
\centering\epsfig{figure=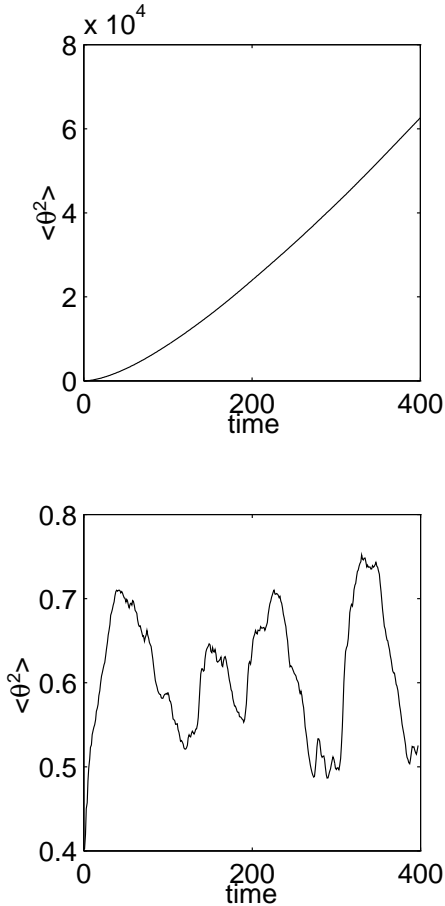,width=6cm}
\begin{minipage}[b]{15cm}
\caption{\baselineskip=12pt
The spatial variance of $\theta$ versus time is plotted for
different temperatures. At the bottom we have
the plot for $T=0.4783$, while on top it is for $T=2.7905$.
 The numerical data is
obtained for a lattice size of $N=256^2$. We clearly see that the
variance of $\theta$ is an increasing function of time only for
$T>T_{KT}$.}
\label{theta(t)}
\end{minipage}
\end{figure}

At higher temperature the weight of random regions increases, and the
size of domains is reduced. In this regime, since the density of
potential energy is bounded, most of the energy is kinetic and
therefore most of the spins are fully rotating, with similar
amplitudes. This is illustrated in Fig.~\ref{timet}, where the
typical temporal behavior of one spin in a high temperature field is
shown. On the contrary, at low temperatures, angles remain bounded in
time. On the other hand the spin momentum takes essentially constant
values (positive or negative ones, depending on the sense of
rotation). The individual behavior of the spins also reflects into
the statistical properties of the system. The angle variance is
bounded at low temperatures, in agreement with the random phase
approximation, while it steadily increases in time at high
temperatures, as we can see in Fig.~\ref{theta(t)}. In analogy with
the properties of a perturbed pendulum, the change of the topology of
trajectories is related to the crossing of a separatrix. Below a
certain value of the energy density $h_c$ ($h_c=h(T_{KT})\approx1$ in
our model), the spins are collectively trapped and after the
transition, as the behavior of the topological defects suggests, they
start to rotate in organized domains, whose size progressively
decreases as the energy density increases.

An important consequence of these results is that the system, at
these temperatures, can be represented by local ordered regions of
synchronized spin motion, and random interfaces where the
defects accumulate. To these rotating spin domains one may add some
level of fast (with respect to the time scale of rotation)
fluctuations characterized by an effective low temperature $\tilde
T$, which takes into account the potential energy. In this context we
may introduce relevant variables $s_i$, which take the values $\pm 1$
according to the sign of spin rotation, $\dot\theta_i$. It is
important to keep in mind that the system is considered to be in its
thermodynamical state, and that the temporal average of spin rotation
velocity $\overline{\dot\theta^2_i}=T$ is, using ergodicity,
independent of the spin site. We can then assume that the angle
velocity is of the form $\dot\theta_i=\Omega_i\approx s_i\sqrt T$,
and write
\begin{equation}
\theta_i = s_i\Omega \;t+\tilde{\theta}_i(t/\epsilon)\;,
\label{tetrot}
\end{equation}
where $\Omega=\sqrt{T}$, and $\epsilon$ characterize the fast
temporal fluctuations of the function $\tilde{\theta}_i$, i.e.
$|\epsilon|\ll1$ and $|\tilde{\theta}_i(t/\epsilon)| \ll
s_i\Omega\,t$ (see Fig.~\ref{timet}). The field $\tilde{\theta}_i$ is
in fact similar to the $\theta_i$ used in (\ref{rand}), but
associated to another temperature $\tilde T$: in a rotating reference
system, attached to each domain (where all the $s_i$'s are equal), it
is an effective ``low temperature'' XY spin field.

The introduction of the new field $s_i=\pm 1$, which label each
lattice site by the sign of the spin rotation, suggests an analogy
between the high temperature XY system and the Ising model. We will
exploit this analogy to construct a model aimed at describing the
complex properties of the system near the Kosterlitz-Thouless
transition, which do not resume to the simple unbinding mechanism of
dipole vortices.

In order to map the XY system into an Ising system, we take advantage
of the time scale separation between the relative slow rotation and
the fast thermal fluctuations. Let us now average the potential
energy over a few rotation periods during which the approximation
made in (\ref{tetrot}) remains valid. For that purpose, we are
reduced to compute the average of terms of the form:
\begin{equation}
\overline{\cos(\theta_i-\theta_j)}=
   \overline{
   \cos[(s_i-s_j)\Omega\,t]\cos(\tilde{\theta}_i-\tilde{\theta}_j)} -
   \overline{
   \sin[(s_i-s_j)\Omega\,t]\sin(\tilde{\theta}_i-\tilde{\theta}_j)}
   \:,
\label{mt}
\end{equation}
where $i$ and $j$ denote two neighbors. The fast temporal behavior of
the $\tilde{\theta}_i$'s assures its fast thermalization at a
temperature $\tilde T$ and decorrelates the two terms in each
product. This allows us to split the time average in two steps, first
on the fast time dependence and then on the slower time scale. The
first average leads to
$$
\overline{\cos(\tilde{\theta}_i-\tilde{\theta}_j)}=
\frac{v(\tilde T)}{2}\;, \qquad
\overline{\sin(\tilde{\theta}_i-\tilde{\theta}_j)}=0\;,
$$
where we use the results of Section \ref{SecAnaly}, due to the
effective low temperature $\tilde T$ within a domain. Then, the mean
of $\cos[(s_i-s_j)\Omega t]$ is zero if the two spins do not rotate
in
the same direction (do not belong to the same domain) and one if the
two spins corotate, which means that the average of this terms can be
represented by a Heaviside function $\Theta(s_i s_j)$. We then obtain
the expression
\begin{equation}
\overline{v}= \frac{1}{2N}
     \sum_{<i,j>}\overline{\cos\left(\theta_i-\theta_j \right)} =
     \frac{v\left(\tilde T\right)}{4N}
         \sum_{<i,j>}\Theta(s_is_j) \:,
\end{equation}
for the mean interaction energy, where the summation is over all
neighbors, and the Heaviside function just states that only
synchronized, corotating spins contribute to the averaged potential
energy. Using the identity $\Theta(s_is_j)=(1+s_is_j)/2$, we finally
obtain a new effective Hamiltonian,
\begin{equation}
H_e = N\left(\frac{T}{2}+2-\frac{v(\tilde T)}{2}\right) -
  \frac{v(\tilde T)}{4}\sum_{<i,j>}s_is_j
  \label{Hequ} \:.
\end{equation}
We notice that this effective Hamiltonian is the sum of two different
terms, which we denote $H_T$ and $H_I$. The first term $H_T$ depends
on the temperature, the size of the lattice and $v(\tilde T)$, and
can therefore be considered as an energy reference term. The second
term $H_I$ introduces a coupling energy between the spins, and can be
recognized as a ferromagnetic Ising-like Hamiltonian whose coupling
constant $J_I(T)=v(\tilde T)/4$ is a function of the temperature.
Using an Ising terminology, we have now a new population of spins
$s_i=\pm 1$, interacting with their close neighbors on a square
lattice. Therefore, under the assumption of ergodicity and taking
into account that spins rotate on a well separated time scale with
respect to their fluctuations, we have mapped the XY system into a
Ising-like model; these two models are linked through a coupling
constant dependent on a temperature related to fluctuations. We shall
note, that the time dependence of the hamiltonian is masked, but the
$s_i$'s are still functions of time, and we still are within the
microcanonical ensemble. Invoking again, the thermodynamical
equilibrium, and the fact that both the microcanonial and canonical
approaches leed to the same thermodynamic limit, we can now continue
our study within the canonical ensemble.

As it is well known, the Ising model undergoes a phase transition
and generate a spontaneous magnetization $M_I$ when $J_I/T >
\beta_{Ic}\approx0.44$ \cite{Landau}. In order to know if the present
Ising system reaches the transition, we have to investigate the
behavior of $v(\tilde T)$. The previously observed domains
(Fig.~\ref{vortpos}), are in agreement with the phenomenology of the
Ising model above its transition, the maximum value of the coupling
constant being $v(\tilde T)/4<1/2$ implies that both models are
surely in their high temperature states for $T>1/2\beta_{Ic}\approx
1.14$. In the low temperature regime, none of the spins are rotating,
$T=\tilde T$ and $v_I=v$, all the potential energy is due to the
waves. We know that the energy is continuous through the
Kosterlitz-Thouless transition and also through the Ising transition.
As a consequence, we expect that $v(T_{KT})=v(\tilde T)$ at the
transition temperature. Moreover, the effective domain temperature
$\tilde T$ has to be an increasing function of $T$, and then, $v_I$
is a decreasing function of $T$, in order to obtain $h=T/2+2$ at high
temperature, $v(\tilde t)$ has to vanish as $T\rightarrow\infty$. The
equation
\begin{equation}
J_I/T=v(\tilde T)/(4T)=\beta_{Ic}
\label{Tc}
\end{equation}
has therefore a unique solution. A calculation of $T_{KT}$, using
equations (\ref{Tc}) and (\ref{vim}) gives $T_{KT}\approx0.855$,
which is in good agreement with the numerical value of
$T_{KT}\approx0.898$ found in the literature.

There is however, an important difference between the usual Ising
model, and the present one, derived from the XY system. It is the
global constraint associated to the conservation of the total
momentum. The mean value of the total momentum $P=0$, is given by
$0=\sum\dot\theta_i=\sum s_i$. But this last expression is precisely
the Ising magnetization:
\begin{equation}
M_I=\frac{1}{N}\sum_is_i=0 \;,
\label{IsingMag}
\end{equation}
which means that only symmetric distribution of positive and negative
spins are allowed. This constraint prevents the system from
undergoing, as in the Ising model, a second order phase transition,
with spontaneous appearence of a macroscopic magnetization below the
critical temperature. In the present case, this would mean that a
significant fraction of the spins would rotate in a preferred
direction, but (\ref{IsingMag}) forbids this kind of phase
transition, and forces the system to accomodate to a vanishing
magnetization. In fact what happens is that, below the transition
temperature, the spins cannot fully rotate, domains dissapear, and
the long range order is established by a reorganization of spin
motions in the form of spin waves, which can still generate a
nonintensive finite-size magnetization.

\section{conclusion}
\label{SecConcl}

In this paper we took a dynamical point of view to analyze the
statistical properties of the XY model. This approach has the
advantage of offering a natural physical framework. From the
structure of the evolution equations of the spins (which become
coupled oscillators in the linear approximation), one is prompted to
consider the phonons (propagating waves in the lattice) as the basic
excitation at low temperature. In the thermodynamic limit, and
assuming that the system reaches an ergodic equilibrium state, it is
reasonable to introduce random phases in the waves. Using this basic
mechanism, we obtained a nonlinear dispersion relation containing the
fundamental physics of the system. Macroscopic quantities, such as
the energy and the (finite size) magnetization, are well described by
this method up to temperatures close to the Kosterlitz-Thouless
critical temperature.

On the other hand, for high temperatures, our dynamical approach
allowed us to map the XY model into an Ising model with a coupling
constant depending on the temperature. This relation to an exact
model, is beneficial to understand the physics near the transition,
and moreover, permits the analytical computation of the critical
temperature in excellent agreement with Monte Carlo simulations.

These results lead to a physical description of the
Kosterlitz-Thouless transition in terms of the change from an ordered
state (long range correlation are established by the wave
excitations) to a local ordered state where macroscopic domains are
separated by interfaces populated with topological defects.
Topological defects have then a tendency to bind themselves in
clusters. The generation of disorder by the unbinding of dipoles, is
accompanied by the fragmentation of the low energy unique domain of
non rotating spins into separated regions of synchronized rotating
spins.

\acknowledgements
We greatfully acknowledge useful discussions with O. Agullo and D.
Benisti. We thank P. C. W. Holdsworth for giving us his Monte Carlo
data used in Figs.~\ref{Fig1-h} and \ref{Fig2-M}. Part of the
numerical simulations were performed at the Centre de Calcul
R\'egional, R\'egion Provence-Alpes-C\^ote d'Azur. A.V. acknowledges
INFM-FORUM for financially supporting his visits to Florence. This
work is also part of the EC network No. ERBCHRX-CT94-0460.

%
%

\end{document}